\documentclass[11pt,twoside]{article}
\usepackage{./asp2014}

\aspSuppressVolSlug
\resetcounters

\begin{document}

\title{Binary Stars in the Orion OB1 Association, Subgroup \textit{a}}
\author{E.~A.~Semenko, I.~I.~Romanyuk, I.~A.~Yakunin, D.~O.~Kudryavtsev, and A.~V.~Moiseeva
\affil{Special Astrophysical Observatory of the Russian Academy of Sciences, Nizhny Arkhyz, Russia; \email{sea@sao.ru}}}


\begin{abstract}
A detailed spectroscopic survey of chemically peculiar (CP) stars in the Orion OB1 association is the most comprehensive observational program in the field of stellar magnetism that has been carried out at SAO so far. As at the end of 2018, we have completed surveying CP stars in the oldest subgroup \textit{a} of the association. In this paper we give a short overview of CP members of the subgroup showing both direct and indirect signatures of multiplicity. Among the overall 11 stars which had been classified as peculiar, we found several good candidates for further detailed study.
\end{abstract}

\section{Association Orion OB1}
In attempting to explore how the stars evolve, we come to the necessity of surveying them in groups like stellar clusters or associations. The main reason for that is in common genesis of the stars and thus in high accuracy of an age evaluation. In the case of hot stars with abnormal chemical composition, which are usually referenced as chemically peculiar or CP stars and form about 15\% fraction of all the stars of comparable temperature~
\citep{2009A&A...498..961R}, the ability to observe the same mass stars of equal age with and without peculiarities becomes extremely important as the global processes are slow. Atomic diffusion is an example of such processes. It takes place in the atmospheres of stars within the wide range of spectral classes where atmospheric layers are stable. The magnetic field is a substantial stabilizing factor, which is believed to be responsible for the phenomenon of magnetic CP stars. These stars permanently host strong magnetic fields that globally cover the stellar surface, and hence they are the perfect test beds to study the influence of the magnetic field on atomic diffusion.

In open clusters, Ap/Bp-stars form a small fraction of stellar population. Inspecting the WEBDA database\footnote{\texttt{https://webda.physics.muni.cz}}, one can note that only few chemically peculiar stars normally occur in a typical cluster. Stellar associations are more populated than individual clusters and the absolute number of stars of interest could reach a dozen or more. This fact explains the importance of the associations for statistical studies. Among the nearest stellar associations, the one in Orion is especially attractive. The celestial position of the Orion OB1 association makes its area completely available for telescopes in both hemispheres. Moreover, as all the parts of Orion OB1 are located in 300-450 pc from the Sun, the vast majority of B-stars there appears brighter than 10 magnitudes and therefore are available for the moderate aperture telescopes.

The oldest part of the Orion OB1 is an 8--12 Myr old subgroup \textit{a}, which is also the closest one among four subgroups selected by \citet{1964ARA&A...2..213B}---the mean distance is order of 360 pc. From 311 members of the Orion OB1a mentioned by~\citet{1994A&A...289..101B} 24 stars have signatures of chemical peculiarities~\citep{2013AstBu..68..300R}. As we are interested only in stars hosting the magnetic field, we excluded from further consideration the non-magnetic members. Among them there are six Am, two HgMn stars, and one emission-line B-star. The rest fifteen stars were selected for observing with the 6-m telescope. A spectropolarimetric survey of CP stars in Orion has been started in 2010. By the end of the observational campaign 2017/2018, we had collected at least three spectra for all potentially magnetic stars in the subgroup \textit{a}. A paper summarizing the observational results, occurrence of magnetic field and spatial distribution of stars is under preparation. In this study, we are examining the radial velocities of CP stars with respect to the possible presence of unknown binary and multiple stars in our sample.

\section{Observational Material}
An observational program of CP stars in subgroup \textit{a} with the Russian 6-m telescope at SAO was accomplished in 2018. This program was carried out in spectropolarimetric mode using the Main Stellar Spectrograph \citep[MSS,][]{2014AstBu..69..339P} installed at the Nasmyth focus of the telescope. The working spectral range, with rare exceptions, was limited to 440--497 nm. In this instance the mean resolving power of spectra $R$ was about 15\,000. The observational technique and data reduction had been demonstrated in multiple papers~\citep[e.g.][]{2006MNRAS.372.1804K, 2017AstBu..72..165R, 2017AstBu..72..384S}.

The total number of spectra were measured in this study is indicated for each object in column $n$ of a summary Table~\ref{table:summary}.

\section{Techniques of Measuring}

Although the region of Orion has been intensively studied for a long time, still there is a lack of information about a multiplicity of Orion's stars. As the mean distance to the association exceeds 350 pc, techniques like speckle-interferometry or adaptive optics lose their efficiency in searches for the new multiple systems. Observing the changes in the stellar radial velocity one can detect the presence of components unseen under other circumstances. Such research was not in our plans, but we decided to check the possible presence of yet unknown spectroscopic binaries in our sample using the available spectra.

The radial velocity can be measured from a displacement of strong spectral lines. In spotted stars, hydrogen lines give more adequate estimates as this element is not suffering from an inhomogeneous surface distribution which is typical for iron, chromium, silicon or magnesium. To measure the radial velocity from hydrogen lines, we fitted the core of the $H_\mathrm{\beta}$ line with a Gaussian function.

For the stars with a significant number of lines in their spectra, we extracted the mean LSD-profiles~\citep[e.g.][]{2010A&A...524A...5K}. Measuring the average $I$ Stokes profiles gives the information about stellar radial velocity and its rotational broadening. All values of the mean projected rotational velocity $v\sin i$ in Table~\ref{table:summary} were estimated from LSD-profiles unless other was specified. Line masks for LSD were constructed upon stellar spectral type. The mean profiles were approximated with a broadening function~\citep{2005oasp.book.....G}.

For measuring the radial or rotational velocity when the line profiles are changing with time, the method of synthetic spectra is a good option. Spectra of those stars in the current study were modeled in the LTE approach; the basic parameters were chosen with respect to spectral classification.

\begin{table}[!ht]
    \caption{Summary of results. New binary and possible binary stars are marked by asterisk in a column ``$RV$''. Column ``$n$'' shows the number of measured spectra.}
    \label{table:summary}
    \smallskip
    \begin{center}
        {\small
            \begin{tabular}{lcccccc}
                \tableline
                \noalign{\smallskip}
                HD & $m_\mathrm{V}$, mag & Sp. type & $\pi$, mas & $v\sin i$ & $RV$, km\,s$^{-1}$  & $n$ \\
                \noalign{\smallskip}
                \tableline
                \noalign{\smallskip}
                33917   &  9.24   &  A0 Si    & 2.61  &  $86\pm10$, $176\pm4$ &  $43.1\pm8.4$, $26.8\pm6.6^{*}$ &  2 \\
                34859   &  9.18   &  A0 Si    & 3.29  &  $85\pm2$            &  $24.1\pm1.2$                   &  4 \\
                35008   &  7.10   &  B9 Si    & 4.87  &  $205$ (syn)         &  $21.9\pm5.9$ (syn)             &  3 \\
                35039   &  4.74   &  B2 He-r  & 2.87  &  $31\pm1$            &  $27.3\pm1.8$                   &  3 \\
                35177   &  8.13   &  B9 Si    & 2.77  &  $220$ (syn)         &  $27.6\pm16.5$                  &  5 \\
                35298   &  7.91   &  B6 He-wk & 2.69  &  $59\pm5$            &  $22.1\pm3.0$                   &  12 \\
                35456   &  6.95   &  B7 He-wk & 0.67  &  $25\pm2$            &  $19.9\pm1.4$                   &  4 \\
                35502   &  7.34   &  B6 SrCrSi& 2.61  &  (Sikora et al. 2016) & (Sikora et al. 2016)           &    \\
                35575   &  6.42   &  B3 He-wk & 2.57  &  $88\pm7$            &  $18.8\pm3.8$                   &  4 \\
                35730   &  7.18   &  B4 He-wk & 2.72  &  $50\pm2$            &  $8.9^{*}$               &  5 \\
                35881   &  7.77   &  B8 He-wk & 2.74  &  $335$ (syn)         &  $22.7\pm11.3$                  &  5 \\
                36429   &  7.56   &  B6 He-wk & 2.67  &  $93\pm2$            &  $27.3\pm1.7$                   &  1 \\
                36549   &  8.54   &  B7 He-wk & 2.75  &  $52\pm5$            &  $20.6^{*}$                     &  8 \\
                38912   &  9.37   &  B8 Si    & 2.13  &  $81\pm5$            &  $12.6\pm4.4$               &  3 \\
                294046  &  8.26   &  B9 Si    & 2.69  &  $129\pm11$ (syn)    &  $26.8\pm11.5$                  &  4 \\
                \noalign{\smallskip}
                \tableline\
            \end{tabular}
        }
    \end{center}
\end{table}

\section{Binary Stars}

Looking into the output of NASA/ADS database, one can note that the interest to binary and multiple systems in Orion is mostly concerned around young stars and protostellar objects. Along with excess of papers devoted to individual stars the number of observational surveys remains very small. Generally, a study by \citet{1991ApJS...75..965M} is the only example of such surveys. Among the 96 most probable members of Orion OB1 that had been studied by the authors, five stars were common in our sample.

\subsection*{Known Systems}
For two stars~--- HD 35008 and HD 35456~--- \citet{1991ApJS...75..965M} had reported variable radial velocities. Hence, in our study these stars are considered as presumably binary systems. The same authors also recognized HD\,35575 as a single-lined binary system and found its orbital elements.

Our observations of HD\,35008 in 2013, 2014, and 2017 showed constant radial velocity $RV=21.9$ km\,s$^{-1}$, though with large scattering. The latter may be caused by the real changes of the radial velocity, but again this scattering might result from substantial spectral variability of the star~(Fig. \ref{figure:hd35008}). After accumulation of all available measurements of $RV$ for this star, we were failed in constructing of its spectroscopic orbit.

\articlefigure[width=.7\textwidth]{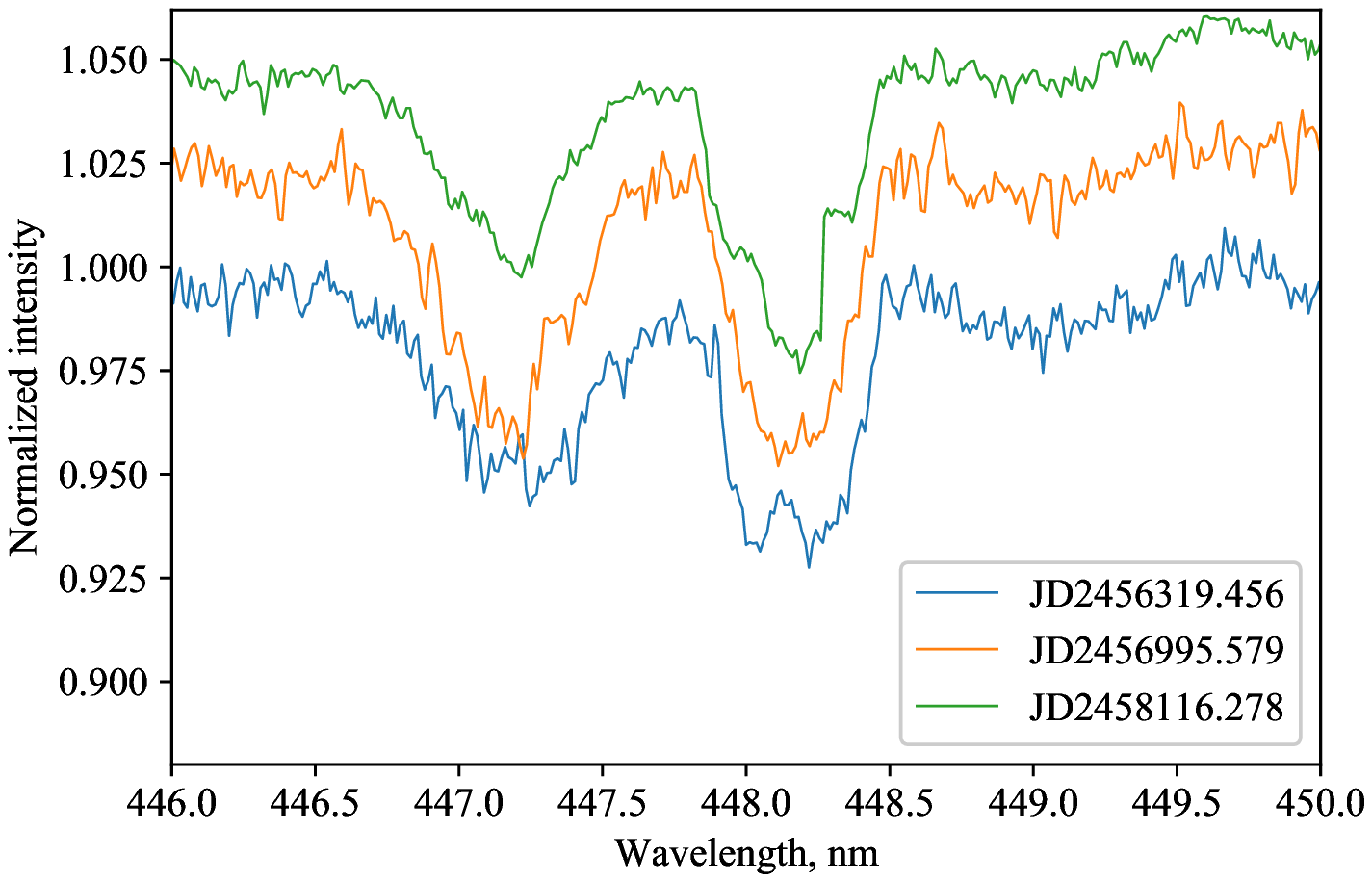}{figure:hd35008}{Spectral variability of He and Mg lines in HD\,35008. Spectra shifted in a vertical direction for demonstration purposes.}

HD\,35456 in our observations had demonstrated a constant radial velocity\linebreak \mbox{$RV=19.9$ km\,s$^{-1}$} with a small scatter. This star had been observed with spectropolarimetry at SAO for five years since 2010. In  Morrell \& Levato's paper~(\citeyear{1991ApJS...75..965M}), the radial velocity of HD\,35456 was declared as variable based on 12 measurements. In particular, during four days $RV$ grew up from 4 to 24 km\,s$^{-1}$. HD\,35456 is a known binary system with a close three magnitudes fainter companion at 0.7 arcsec away~\citep{2012AstBu..67...44B}.

Conclusion about multiplicity of HD\,35575 made by Morrell \& Levato relies generally on their own observations which had been made during a week in 1980. Proposed orbital solution suggests very short 2.24 days period and an eccentricity $e=0.16$. Neither of our four observations agrees with published orbital parameters, though we confirm a moderately changed radial velocity of the star~(Fig. \ref{figure:hd35575}, Table~\ref{table:hd35575}).

\begin{table}[!ht]
    \caption{Radial velocity of HD\,35575.}
    \label{table:hd35575}
    \smallskip
    \begin{center}
        {\small
            \begin{tabular}{l l}
                \tableline
                \noalign{\smallskip}
                HJD, 2450000+ & $RV$, km\,s$^{-1}$  \\
                \noalign{\smallskip}
                \tableline
                \noalign{\smallskip}
                5553.309  &  $23.7\pm4.4$                \\
                7830.299  &  $20.4\phantom{\,\pm\,}4.2$  \\
                8008.586  &  $13.5\phantom{\,\pm\,}4.4$  \\
                8068.524  &  $17.5\phantom{\,\pm\,}4.5$  \\
                \noalign{\smallskip}
                \tableline\
            \end{tabular}
        }
    \end{center}
\end{table}

\articlefigure[width=.7\textwidth]{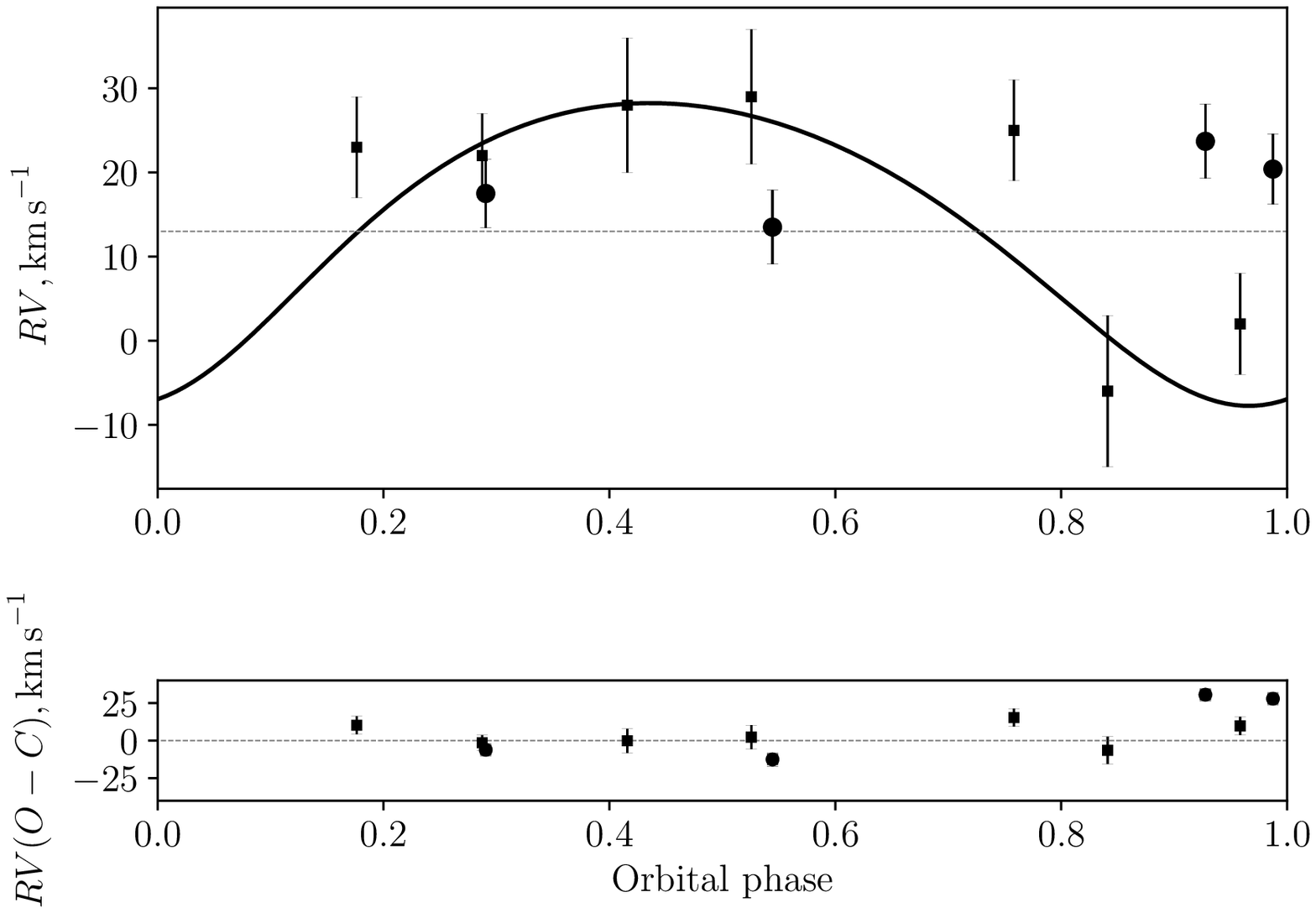}{figure:hd35575}{Radial velocity of HD\,35575 phased according to ephemerides of binary system published by \citet{1991ApJS...75..965M}. Measurements from Table~\ref{table:hd35575} are displayed as filled dots.}

The last star in the list of known binary and multiple stars is HD\,35502. This massive magnetic CP star is a member of a triple system. An exhaustive study of this system was done by \citet{2016MNRAS.460.1811S}.

\subsection*{New Binary and Possible Binary Stars}

Variable radial velocity or a presence of two sets of lines in a composite spectrum were the reason to classify the stars HD\,33917, HD\,35730, and HD\,36549 as the new spectroscopic binaries.

\textsl{HD\,33917}. The mean profiles extracted from intensity spectra of HD\,33917 have a complex triangular shape. From the best fitting solution, we have concluded that HD\,33917 is a possible SB2 system consisting of two stars which though have the similar radial velocities at the moments of observation. This pattern is repeating in all observations. A sample of the mean LSD-profile is presented in Fig.~\ref{figure:hd33917}. In observations, the radial velocity of broader component remains nearly constant with small changes~(see Table~\ref{table:hd33917}). The measured projected rotational velocity averaged over four spectra in this case is $\langle v\sin i \rangle _2=176.2\pm3.8$\,km\,s$^{-1}$. A narrow-lined component rotates slower, its projected velocity $\langle v\sin i \rangle _1=86.1\pm9.9$\,km\,s$^{-1}$. Significant spectral variability and a variable radial velocity may lead to the higher error in results.

\articlefigure[width=.7\textwidth]{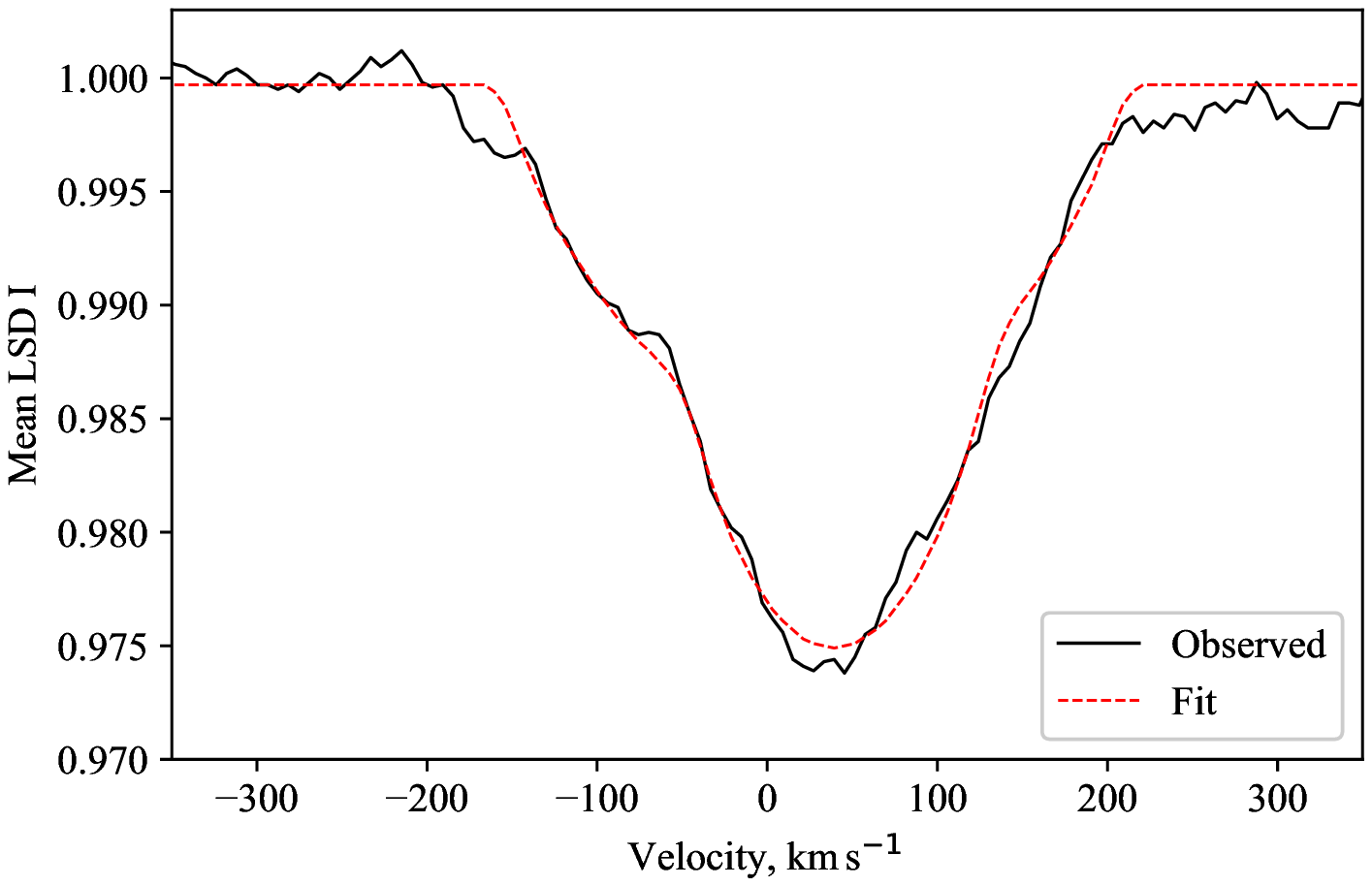}{figure:hd33917}{Mean LSD intensity profile extracted from a spectrum of HD\,33917.}

\begin{table}[!ht]
    \caption{Radial velocity of HD\,33917.}
    \label{table:hd33917}
    \smallskip
    \begin{center}
        {\small\begin{tabular}{l l l}
                \tableline
                \noalign{\smallskip}
                HJD, 2450000+ & $RV_1$, km\,s$^{-1}$ & $RV_2$, km\,s$^{-1}$ \\
                \noalign{\smallskip}
                \tableline
                \noalign{\smallskip}
                6589.461  &  $43.6\pm6.8$                &  $27.3\pm5.9$               \\
                6640.365  &  $31.4\phantom{\,\pm\,}7.1$  &  $34.2\phantom{\,\pm\,}4.9$ \\
                8116.200  &  $36.7\phantom{\,\pm\,}8.4$  &  $23.4\phantom{\,\pm\,}5.0$ \\
                8125.339  &  $60.8\phantom{\,\pm\,}10.6$ &  $22.4\phantom{\,\pm\,}9.4$ \\
                \noalign{\smallskip}
                \tableline\
            \end{tabular}}
    \end{center}
\end{table}

Trying to find the other indicators of stellar multiplicity, we have analyzed photometric series collected in the All-Sky Automated Survey~(ASAS, \cite{1997AcA....47..467P}). Observations of HD\,33917 had been lasting for eight years and thus allowed to find long trends in data. The longest reliable period that could be derived from ASAS photometry is 2654 days, which is comparable with the length of time series. The amplitude of variability is an order of 10 mmag. Bearing in mind these parameters, we cannot exclude an instrumental origin of long term photometric variability in ASAS data, although the long period agrees well with the observed amplitude of stellar radial velocity.

\textsl{HD\,35730}. This star had already been observed spectroscopically by~\citet{1991ApJS...75..965M}. Six individual observations night by night did not show any changes in stellar radial velocity, it stayed constant around 29 km$\,$s$^{-1}$. We have been observing HD\,35730 five times from 2010 till 2017. Our results in Table~\ref{table:hd35730}
witness the radial velocity reversing its sign. Apparently, HD\,35730 is a new single-lined spectroscopic binary.

\begin{table}[!ht]
    \caption{Radial velocity of HD\,35730.}
    \label{table:hd35730}
    \smallskip
    \begin{center}
        {\small\begin{tabular}{l l l}
                \tableline
                \noalign{\smallskip}
                HJD, 2450000+ & $RV$, km\,s$^{-1}$ \\
                \noalign{\smallskip}
                \tableline
                \noalign{\smallskip}
                5553.244  &  $18.8\pm2.5$               \\
                5841.585  &  $-5.4\phantom{\,\pm\,}2.6$ \\
                5842.572  &  $-5.6\phantom{\,\pm\,}2.7$ \\
                5962.285  &  $22.7\phantom{\,\pm\,}2.6$ \\
                7740.350  &  $13.7\phantom{\,\pm\,}2.6$ \\
                \noalign{\smallskip}
                \tableline\
        \end{tabular}}
    \end{center}
\end{table}

Small amount of acquired measurements does not suggest finding of optimal orbital solution at this time. However, we can select one possible solution with a period $P_\mathrm{orb}=909.86$ days and nearly circular orbit. The complete set of orbital parameters is in Table~\ref{table:hd35730orb}. To obtain this solution we use after averaging the original measurements from a paper of~\citet{1991ApJS...75..965M}. In the future we plan to explore the star with new measurements.

\begin{table}[!ht]
    \caption{Orbital parameters of HD\,35730 in the case of a long orbital period.}
    \label{table:hd35730orb}
    \smallskip
    \begin{center}
        {\small\begin{tabular}{l l}
                \tableline
                \noalign{\smallskip}
                Parameter & Value \\
                \noalign{\smallskip}
                \tableline
                \noalign{\smallskip}
                $P$ (d)                 &  $909.86\pm1.63$         \\
                $T_\mathrm{p}$ (HJD)    &  $2455868.064\pm36.684$    \\
                $e$                     &  $0.0493\pm0.0869$         \\
                $\omega$ (deg)          &  $225.6\pm14.5$            \\
                $\gamma$ (km$\,$s$^{-1}$) & $24.8\pm3.2$             \\
                $K_{1}$ (km$\,$s$^{-1}$)  & $37.7\pm6.9$             \\
                \noalign{\smallskip}
                \tableline\
        \end{tabular}}
    \end{center}
\end{table}

The radial velocity curve computed for the presented orbital solution is plotted in Fig.~\ref{figure:hd35730orb}.

\articlefigure[width=.7\textwidth]{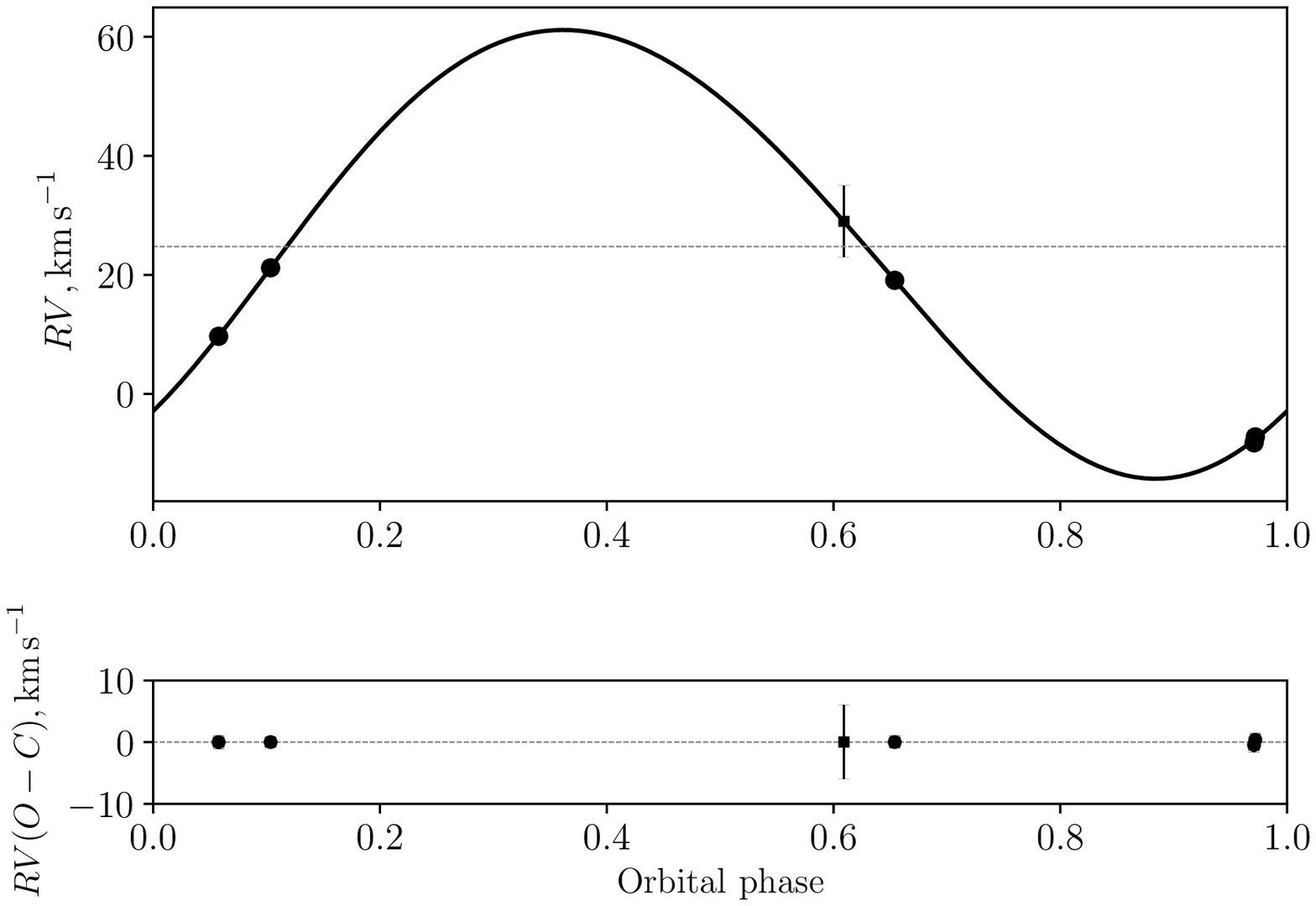}{figure:hd35730orb}{Radial velocity of HD\,35730 according to a solution with a circular orbit and long period. Different symbols mean the same as in Fig.~\ref{figure:hd35575}.}

\textsl{HD\,36549}. This star has been observed at SAO for five years. In eight spectra we did not find any signatures of a magnetic field, but the other attributes of CP stars like a spectral variability and abnormal intensity of lines were clearly seen.

On six unevenly spaced nights, the measured radial velocity of HD\,36549 ranged from $-7.4$ to 51.9 km$\,$s$^{-1}$~(Table~\ref{table:hd36549}). At two moments separated by three years, the measurements gave exactly the same value of $RV$. All together, these facts let us to put limits on the orbital parameters of the system. In our vision, HD\,36549 is a single-lined binary system with a nearly circular orbit ($e=0$--0.1) and an orbital period $P_\mathrm{orb}$ which is a factor of 1095.92 days. At the moment, the available number of measurements is insufficient to make the final decision about the correct value of $P_\mathrm{orb}$.

\begin{table}[!ht]
    \caption{Radial velocity of HD\,36549.}
    \label{table:hd36549}
    \smallskip
    \begin{center}
        {\small\begin{tabular}{l l l}
                \tableline
                \noalign{\smallskip}
                HJD, 2450000+ & $
                RV$, km\,s$^{-1}$ \\
                \noalign{\smallskip}
                \tableline
                \noalign{\smallskip}
                6224.560  &  $51.9\pm3.3$               \\
                6639.536  &  $38.9\phantom{\,\pm\,}3.1$ \\
                6639.558  &  $39.3\phantom{\,\pm\,}3.0$ \\
                6644.466  &  $26.0\phantom{\,\pm\,}3.2$ \\
                7740.381  &  $26.0\phantom{\,\pm\,}3.3$ \\
                7763.245  &  $-4.1\phantom{\,\pm\,}2.9$ \\
                7763.267  &  $-5.8\phantom{\,\pm\,}3.0$ \\
                8068.544  &  $-7.4\phantom{\,\pm\,}2.9$ \\
                \noalign{\smallskip}
                \tableline\
        \end{tabular}}
    \end{center}
\end{table}

\section{Rotational Velocities}

A significant fraction of stellar population in associations consists of the early B and O-stars which typically rotate fast. However, in our sample the fastest rotators are among late B-stars with He-wk or Si anomalies. When the projected velocity $v\sin i$ approaches to 200 km$\,$s$^{-1}$ the modeling becomes needed to measure stellar rotation. This could be a tricky problem because the number of lines decreases as the effective temperature goes higher. That is why we used the different methods for slowly or moderately rotating stars and for those with $v\sin > 200$ km$\,$s$^{-1}$.

The rotational velocity of HD\,35008 was found from the best fit of Mg and He lines. At some moment, the lines of He\,\textsc{i} 4471\,\AA\ and Mg\,\textsc{ii} 4481\,\AA\ showed the structure usually seen in stars with spotted surface~\ref{figure:hd35008}. HD\,35177 and HD\,294046 generally follows the same behavior. The measured rotational velocity of HD\,35008 and HD\,35177 exceeds 200\,km\,s$^{-1}$, HD\,294046 rotates slower~($v\sin i=129$ km\,s$^{-1}$). The outstanding object in our sample is HD\,35881, a late B-star with the rotational velocity $v\sin i=335$\,km\,s$^{-1}$. For the rest stars the rotational velocities are well distributed within 20--100 km\,s$^{-1}$ with a slow trend to higher values.

\section{Conclusions}
In the current study we have analyzed a sample of chemically peculiar members of the Orion OB1 association, subgroup \textit{a}. The conclusion about star's membership relies on a fundamental paper by \citet{1994A&A...289..101B}. The measured stellar radial velocities generally confirm that the stars belong to Orion OB1 as their values are close to 23km\,s$^{-1}$, the mean radial velocity of the association itself. Also, ignoring the difference in $RV$ for some newly discovered spectroscopic binary systems, their systemic velocity is agreed well with other association members.

Our sample contains only potentially magnetic CP stars, excluding HgMn and Am. The presence of slow rotators without measurable magnetic field in the sample nonetheless make possible the wrong classification and thus the stars HD\,35730 and HD\,36549 might be in fact HgMn.

Before this work was started, only four binary or multiple stars out of 15 in the sample had been known. We have shown that at least three another stars have a variable radial velocity that allows to classify them as new spectroscopic binaries. So far, the occurrence of binaries among potentially magnetic CP stars in Orion OB1a becomes 46\%. The observational program aimed at a deep study of Orion OB1 is ongoing and the new results will appear. In our plans, there is an abundance analysis which should clarify the type of chemical peculiarities of selected stars in Orion.

\acknowledgements The authors acknowledge the Russian Foundation
for Basic Research for partial financial support of this
study~(grant No.18-52-06004 Az\_a). This research has made use of
the WEBDA database, operated at the Department of Theoretical
Physics and Astrophysics of the Masaryk University. Observations
at the SAO RAS telescopes are supported by the Ministry of Science
and Higher Education of the Russian Federation.

\bibliography{semenko}  

\end{document}